\documentclass[3p,final,11pt,authoryear]{elsarticle}
\usepackage{subcaption}
\usepackage{rotating}
\usepackage{tikz}
\usepackage{booktabs}
\usepackage{graphicx}
\usepackage{amssymb}
\usepackage{natbib}
\usepackage{amsthm}
\usepackage{multirow}
\usepackage{amsmath}
\usepackage[linesnumbered,ruled]{algorithm2e}
\newsavebox{\measurebox}
\usepackage{lineno}
\usepackage{array,calc}
\usepackage{lscape}
\usepackage{float}
\usepackage{hyperref}
\usepackage[nameinlink,capitalize]{cleveref}
\usepackage{subcaption}

\floatstyle{plaintop}
\restylefloat{table}

\usepackage{xcolor}
\usepackage{adjustbox}
\usepackage{tabularx}
\usepackage[symbol,multiple]{footmisc}
\usepackage{array,calc}
\usepackage{makecell}

\usepackage{natbib}

\journal{}

\begin{document}
\begin{frontmatter}
\title{XARA: A Multimodal Agentic System for eXtended Reality Experiment Generation}
\author[1]{Tareq Alsaleh}
\author[1]{Thomas Zhao}
\author[1]{Luke Guardino}
\author[1]{Keagan Valentine}
\author[1]{Isabel Kim}
\author[1]{Bilal Farooq \footnote{Corresponding author (bilal.farooq@torontomu.ca)}}

\address[1]{Laboratory of Innovations in Transportation (LiTrans), Toronto Metropolitan University, Toronto, Canada}

\begin{abstract}
Controlled experiments are essential for advancing transportation research, yet designing eXtended Reality (XR) environments for behavioural studies remains technically demanding and inaccessible to many domain researchers. Current workflows require proficiency in game engines such as Unity or Unreal Engine, imposing significant barriers to rapid scenario prototyping and iterative experiment design. This paper presents XARA, a novel system that integrates a dual large language model (LLM) architecture with an immersive XR environment to enable voice driven experiment generation for transportation research. XARA employs a pipeline in which spoken natural language commands are first transcribed and then interpreted into structured commands that manipulate a Unity based XR scene in real time. The system supports object creation, spatial manipulation, animation control, and scene persistence through a modular architecture. By lowering the technical threshold for XR experiment creation, XARA enables transportation researchers to rapidly prototype driving, pedestrian, and urban design scenarios using intuitive voice commands and hand gestures within an immersive 3D environment. This paper describes the system architecture, the voice to command pipeline, and discusses potential applications across behavioural travel experiments, safety studies, human factors research, and policy testing.

\end{abstract}

\end{frontmatter}
\section{Introduction}
\label{S:TInt}
Controlled experiments form the backbone of empirical transportation research. Understanding how travellers make decisions, how pedestrians interact with vehicles, and how infrastructure design shapes behaviour requires carefully constructed experimental environments where variables can be isolated and manipulated \citep{fisher2011handbook, BENELIA2010249, pedestrian_vr}. Over the past two decades, virtual reality (VR) has emerged as a powerful tool for conducting such experiments, offering a combination of ecological validity, experimental control, and participant safety that is difficult to achieve in field studies.

Despite these advantages, the adoption of XR in transportation research has been constrained by significant technical barriers \citep{thomsen2026opensource}. Constructing a XR experiment typically requires proficiency in game engine development environments such as Unity or Unreal Engine, demanding skills in 3D modelling, C\# or C++ programming, physics simulation, and user interface design. For many transportation researchers whose expertise lies in econometrics, traffic engineering, or behavioural science rather than software development, these requirements represent a prohibitive obstacle. The result is that XR experiment creation often depends on dedicated technical staff or external collaborators, leading to long iteration cycles and limited researcher autonomy over scenario design.

Recent advances in large language models have introduced new possibilities for bridging the gap between natural language intent and complex system operations. In the past few years, highly complex AI models, such as those from OpenAI, Anthropic, and Google, and their successors have demonstrated remarkable capabilities in understanding nuanced instructions, generating structured outputs, and reasoning about spatial and procedural tasks \citep{bubeck2023sparks}. Simultaneously, automatic speech recognition systems have achieved near-human accuracy in transcribing spoken language across diverse accents and acoustic conditions \citep{radford2023robust}.

This paper presents XARA, an agentic system that leverages these advances to enable voice, and motion XR experiment generation for transportation research. XARA integrates a dual-LLM pipeline, comprising a voice model for speech recognition and a language model for command interpretation, with a Unity-based XR environment equipped with OpenXR support for immersive headset interaction. Researchers can create, move, rotate, scale, and delete scene objects using natural spoken commands and hand gestures, with the system translating utterances into structured JSON instructions that are executed in real time within the XR scene.

The core contributions of this work are threefold. First, we present a novel dual-LLM architecture that separates acoustic understanding from semantic command extraction, providing robustness and modularity. Second, we describe a complete voice-to-scene pipeline that bridges language processing with Unity-based XR rendering through inter-process communication. Third, we discuss the implications of this approach for democratizing XR experiment design in transportation research, enabling rapid scenario prototyping by non-technical domain experts.

\section{Background}
\label{S:tBack}
This section reviews the current state of immersive travel behavioural studies, identifies the significant technical barriers, and emerging technologies that can be adopted to address the barriers.

\subsection{XR in Transportation and Behavioural Research}

Extended reality (XR) technologies are an encompassing family of technologies that includes augmented reality (AR) and virtual reality (VR), and can be described as interactive systems that replace or augment a user’s sensory input with computer-generated stimuli. By continuously tracking the user’s position and actions and updating the virtual environment in response, these systems can create a convincing sense of presence within an artificially generated environment \citep{LaValle_2023,slater1997framework}. In recent years, XR has been increasingly adopted within the transportation and behavioural research community \citep{farooq2018virtual, wong2024virtual, farooq2024workshop,vr_travel_survey_workshop}. Due to its ability to simulate highly immersive and dynamic environments, XR enables researchers to create scenarios for research that would otherwise be difficult or impossible to reproduce with traditional research methods. By using XR as a tool to collect data, participants are able to experience and interact with these scenarios in real time, allowing natural behavioural responses in situations requiring complex econometric trade-offs and spatial perception.

This leads to XR's first inherent advantage in comparison to traditional data collection methods; its ability to provide an intuitive method to study complex decision making scenarios. A popular example in the transportation community is in studying commuting choice \citep{zhao_isctsc}. While traditional survey methods like web-based stated preference questionnaires have frequently been used to determine this question, commuting preferences are highly context based. It is not sufficient to simply present choices for commuting, if the participant relies on their own subjective mental recreation of the environment they make the choice in, as this induces hypothetical bias \citep{HENSHER2010_hypo_bias}. Such survey methods also do not account for additional environmental factors. Simulating neighbourhoods for commuting choice decisions in XR can allow participants to naturally evaluate factors within the scene, like safety, weather, and crowdedness. Because observed behaviour is often shaped by these factors and their complex interactions, traditional survey approaches provide only a partial interpretation of true preferences \citep{ben1985discrete}. XR addresses the previous issues of relying on subjective recreations of environments. In a direct comparison to text-only surveys, XR was proven to address limitations in traditional surveys through the intake of important audiovisual cues \citep{zachary_comparing_text_and_vr}. XR thus allows researchers to present a participant with realistic approximations of the target environment, which a participant experiences directly while making decisions. This reduces the hypothetical bias inherent in stated preference surveys, while retaining the experimental control that revealed preference surveys lack.

XR is also valuable for simulating hypothetical or emerging mobility scenarios. Researchers are able to study situations that would otherwise be impractical, dangerous, or too expensive to consider. For example, interactions in pedestrian crossing research where participants are asked to cross a street while automated vehicles are moving on the street \citep{pedestrian_vr}. Another case is in understanding driver behaviour as XR can utilize motion rigs to produce realistic driving scenarios, such as changing lanes on a highway, navigating through a busy urban environment, or interacting with the automation in Level 3 vehicles \citep{ansar2026analyzing}. 

Despite these advantages, widespread adoption of XR within transportation research currently remains a challenge \citep{thomsen2026opensource}. There are substantial technical and logistical barriers preventing many parties who are interested. XR experiments will usually require highly specialized, and often expensive, hardware such as head-mounted displays like XR headsets, that must be worn to view the immersive environment. The environment itself requires high-performance rendering systems, which are especially costly and necessary to ensure that performance and realism are both sufficient. Addressing any operational inefficiencies would require a deep understanding of computer graphics optimization methods such as hybrid lighting, dynamic LOD systems, and running the experiment on high-end GPU hardware \citep{hybrid_lighting_vr, xia1997_dynamic_lod}. If the researcher wants to integrate other supplementary data like eye tracking or electrodermal activity, it usually requires integration of XR with other third party devices or a specialized VR headset, further complicating the logistics of performing an experiment \citep{eye_tracking, pedestrian_vr}.
Furthermore, the development of these immersive environments are typically conducted in game engines like Unity or Unreal Engine. Developing an XR scene requires specialized expertise in using these game engines, along with understanding user interaction design, hardware configuration, graphic optimization, and other niche understandings. The skills and knowledge required are rarely found in those working in transportation research. This forces research teams to either invest significant time into training people, usually students, or outsourcing development to expensive external XR developers. 

Importantly, the challenge extends beyond XR development itself. XR authoring is fundamentally a workflow problem. Researchers must translate conceptual experimental ideas into executable immersive environments through multiple disconnected layers of tooling and implementation. Spatial layouts, object behaviours, environmental logic, interaction systems, and interface controls are often configured separately through specialized editor interfaces or code-based pipelines. Consequently, even small modifications to experimental conditions may require disproportionate redevelopment effort.

This disconnect between conceptual experiment design and technical implementation has shaped the evolution of XR authoring systems over the past decade. Earlier approaches attempted to reduce the complexity of immersive creation through immersive manipulation, rapid prototyping, collaborative workflows, and low-code interaction systems. More recently, large language models (LLMs) have begun to be explored as tools for behavioural and experimental research, including the simulation of human responses and the replication of established human-subject experiments \citep{aher2023using}. Their use has similarly expanded into transportation research across areas such as traffic analysis, simulation, autonomous systems, and travel behaviour \citep{nie2025exploring,yan2025large, liu2024can, alsaleh2025towards,alsaleh2025generative}. These developments create an opportunity to extend natural-language interaction to XR authoring, allowing researchers to translate experimental intentions directly into modifications of immersive environments.

\subsection{Assisted XR Workflows}

Early XR research recognized that immersive environments should ideally be designed within the same medium in which they are ultimately experienced. Systems such as immersive authoring and WYXIWYG (``What You eXperience Is What You Get'') proposed that users should manipulate virtual content directly inside immersive environments rather than relying exclusively on desktop-based interfaces \citep{lee2004immersive,lee2005immersive}. These approaches reduced the cognitive gap between scene specification and experiential evaluation by allowing users to build and assess virtual worlds simultaneously.

Subsequent work focused on reducing friction during rapid XR prototyping. ProtoAR introduced lightweight physical-digital workflows for prototyping AR applications \citep{nebeling2018protoar}, while Pronto supported AR experience construction through sketches, enacted behaviour, and storyboard-style authoring \citep{leiva2020pronto}. GestureWiz further explored human-in-the-loop interaction prototyping, allowing designers to iteratively test and refine gesture-based interfaces before implementing full production systems \citep{speicher2018gesturewiz}. Collectively, these systems demonstrated that XR prototyping becomes substantially more accessible when creators can externalize ideas quickly without immediately committing to complex implementation pipelines.

In more recent years, advances in Generative Artificial Intelligence (AI) have further expanded the accessibility of XR content creation by enabling users to describe environments through natural language and multimodal interactions. As these systems have evolved, researchers have identified several expectations that users hold when interacting with AI-powered XR agents. Through an investigation of embodied prompting in immersive environments, researchers found that users expect AI agents to possess an understanding of the surrounding environment, interpret embodied inputs such as gestures and spatial references, maintain awareness of previous scene states, and apply common-sense reasoning about objects and their relationships \citep{how_ppl_prompt_gen_ai}. These findings highlight that effective XR creation tools require more than language understanding alone; they must also reason about spatial context and user intent within immersive environments.

Building upon these insights, several systems have explored the integration of large language models into XR authoring workflows. VRCopilot introduced a multimodal co-creation framework that combines natural language, gestures, and direct manipulation to assist users in constructing virtual environments \citep{vr_copilot}. Their findings suggest that while AI assistance can accelerate scene creation, workflows that preserve user control through manual specification provide the greatest sense of agency and ownership over the resulting environment. Other research has investigated higher levels of automation. Holodeck demonstrated that large language models can be also used to generate virtual environments from textual descriptions \citep{holodeck}. Rather than focusing on individual object placement, the system enables rapid generation of entire virtual worlds. Similarly, LLMR explored how language models can function as interactive agents within mixed reality environments, supporting contextual reasoning and task execution through conversational interaction \citep{llmr}. Together, these systems illustrate a broader shift from traditional XR authoring tools toward AI-assisted and AI-generated workflows, where users increasingly collaborate with intelligent agents to create, modify, and navigate immersive environments.

While these approaches significantly reduce the technical barriers associated with XR content creation, they generally focus on either fully automated scene generation or AI-assisted design within a single platform. XARA builds upon this body of work by emphasizing researcher-driven XR authoring through manual multimodal prompting while introducing a standardized workflow that integrates not only natural language but gestures, scene management, and XR-specific design considerations into a unified framework for transportation and behavioural research applications.

\section{Methodology}
\label{Methodology}
This section describes the design and implementation of the XARA system. It first presents the overall system architecture, followed by the voice command processing pipeline and the LLM-based command interpretation. It then describes the controller-based interaction subsystem, including ray-casting and object selection, manual object manipulation, menu navigation, and prefab placement within the XR environment. The scene generation and manipulation subsystem is subsequently presented, detailing how voice- and controller-based inputs are translated into modifications of the virtual environment. The methodology then introduces the user experience and orientation features developed to support interaction with and learning of XARA, before concluding with the asset generation and manipulation framework for integrating generative 3D assets into the system.

\subsection{System Architecture}
\label{system}
The core XARA command-processing architecture consists of four interconnected layers: the Input Layer, the Processing Layer, the Command Layer, and the Execution Layer. These layers support the translation of user inputs into executable modifications of the XR environment, while additional controller-based interaction and user-interface components operate directly within Unity.

Currently, the Input Layer encompasses two input modalities. The primary modality is voice input captured through the system microphone using Unity's \texttt{Microphone} class at 16,000 Hz mono sampling. The secondary modality is XR controller input, which provides ray-based object selection, scaling, rotation, relocation, menu interaction, and a toggle control for activating and deactivating voice processing.

The Processing Layer implements the dual-LLM pipeline accessed via API calls. The first language model, is a voice based model, for example: \texttt{whisper-large-v3-turbo}, which performs the automatic speech recognition, converting raw audio waveforms into text transcripts. The second language model, is a general knowledge model, for example: GPT-4.1, which performs natural language understanding, converting free-form text transcripts into structured JSON command objects. The system is also equipped with a heuristic fallback parser to provide execution alternatives in cases where the LLM is unavailable or returns malformed output.

The Command Layer defines the structured interface between natural language processing and scene manipulation. Commands are represented as JSON objects with three fields: \texttt{action}, \texttt{object}, and \texttt{params}. The \texttt{action} field specifies the operation to perform, for instance move, rotate, add, etc. The \texttt{object} field identifies the target entity, for example: street lamp, pedestrian, car, etc. Finally, the \texttt{params} field contains operation-specific parameters such as coordinates, angles, or scale factors. This schema provides a minimal yet sufficient vocabulary for the supported scene operations.

The Execution Layer operates within the Unity game loop. Incoming commands are first placed in a thread-safe command queue, which acts as a controlled waiting line for storing commands until Unity is ready to process them safely and in order. During each frame update, the main Unity thread retrieves the next command and converts the JSON from a text-based command into an internal command object. The command is then dispatched to the appropriate handler method for execution in the XR scene. The handler then modifies the XR scene through Unity's scene-manipulation APIs, which are the built-in programming interfaces used to change objects within a Unity scene. Through these APIs, the system can update an object's position, rotation, and scale, or create and place new objects in the XR environment when an object-generation command is received.
The complete pipeline can be summarized as follows:

\begin{quote}
User Voice $\longrightarrow$ Microphone Capture $\longrightarrow$ Voice Model Automatic Speech Recognition (ASR) $\longrightarrow$ Text Transcript $\longrightarrow$  General Language Model for Natural Language Understanding $\longrightarrow$ Output JSON Command $\longrightarrow$ Inter-process Communication $\longrightarrow$ Unity Queue $\longrightarrow$ Command Dispatch $\longrightarrow$ Scene Modification $\longrightarrow$ XR Environment Update
\end{quote}

\begin{figure}[htp]
    \centering
    \includegraphics[width=0.85\textwidth]{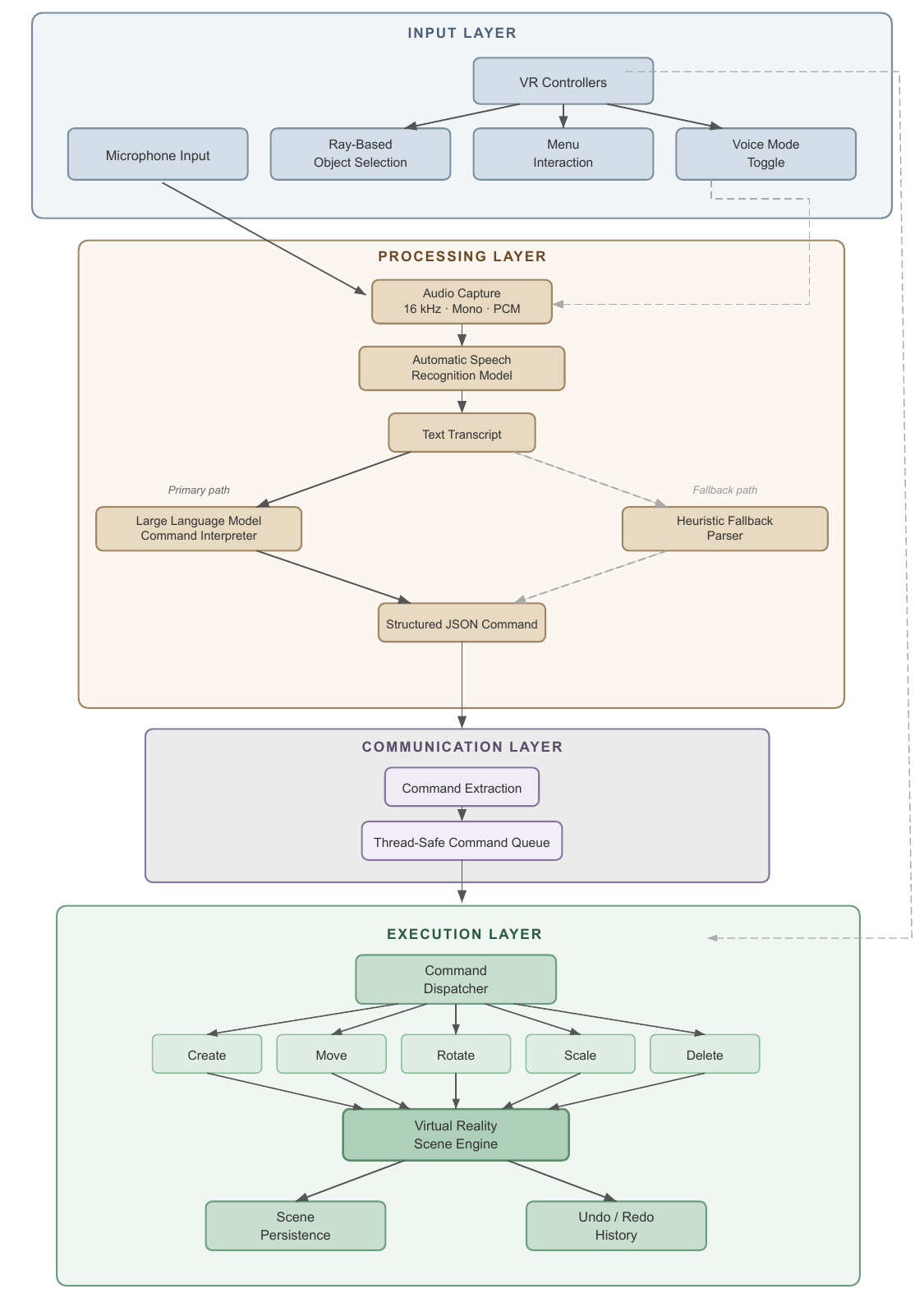}
    \caption{System Architecture of XARA Implementation Pipeline}
    \label{fig:system_architecture}
\end{figure}

Figure \ref{fig:system_architecture} visualizes the system architecture of the implementation pipeline.   

\subsection{Voice Command Processing}
\label{voice}
The voice command processing is done with C\# in Unity. It operates with push to talk, so researchers hold down the voice activation button to enable the microphone, they then utter their command, and once they are done they release the voice activation button. Once they release it, this indicates to the system that the command is complete and it passes the complete recording along to the Whisper model for transcription. We pass this through a Unity WebRequest. 

Audio capture is performed within Unity at 16,000 Hz mono recording with 16-bit integer pulse code modulation (PCM) encoding. A sampling rate of 16,000 Hz is widely used in speech recognition as it captures the frequency range of human speech while keeping data size and processing requirements moderate. Mono recording is sufficient for voice input and avoids unnecessary duplication of channels. The 16-bit resolution provides adequate dynamic range for clear speech capture without excessive storage overhead. PCM here refers to a standard method of representing audio as raw, uncompressed digital samples, which preserves signal fidelity and ensures compatibility with speech recognition models.

Upon utterance completion, as indicated by release of the push to talk button, the audio data is written to a WAV buffer and transmitted using a Unity WebRequest to the speech transcription model for transcription. The API call sends raw WAV bytes with Bearer token authentication to the model endpoint. The response contains the text transcription of the spoken utterance.

The transcript is then preprocessed and automatically formatted in a prompt template for the natural language model. The system prompt instructs the model to return only valid JSON with the keys \texttt{action}, \texttt{object}, and \texttt{params}, ensuring deterministic structured output. The temperature parameter, which controls the model's stochastic behaviour, is set to 0 to maximize output consistency. 

If the LLM call fails because of network errors, timeouts, or rate limiting, the system falls back to a heuristic parser. This parser uses rule-based extraction: the first word is mapped to an action, the next non-article word is mapped to an object, and numeric values are extracted using regular expressions. Action and object aliases are resolved through predefined dictionaries with fuzzy matching with a 0.82 similarity score cutoff.

\subsection{LLM Command Interpretation}
\label{command}

The dual-LLM architecture separates speech recognition and command interpretation across two specialized models, each optimized for its respective function.
The first LLM, voice model, OpenAI Whisper model was used which is a sequence-to-sequence transformer model trained on 680,000 hours of multilingual audio data~\citep{radford2023robust}. Its role is acoustic: converting audio waveforms into text. The turbo variant used in XARA provides reduced latency compared with the full Whisper large model while maintaining high transcription accuracy. By delegating speech recognition to a dedicated model, the system achieves robustness to accent variation, background noise, and acoustic conditions typical of XR laboratory settings.

The second LLM, the natural language processing model, GPT-4.1, performs the semantic interpretation of the transcribed text \citep{achiam2023gpt}. This model receives the raw transcript together with a system prompt that defines the expected output schema and the available action vocabulary. The model maps diverse natural language phrasings to a fixed set of canonical actions and extracts relevant parameters. For example, the utterances ``put the cube over at five ten fifteen'', ``move cube to coordinates 5, 10, 15'', and ``shift the box to position five ten fifteen'' should all produce the same structured output.

The JSON command schema supports the following structure:

\begin{verbatim}
{
  "action": "move | rotate | resize | add | delete | undo/ redo | filter",
  "object": "string",
  "params": {
    "x": 0.0
  }
}
\end{verbatim}

The system maintains alias mappings for both actions and objects to handle synonyms. Action aliases include mappings such as \texttt{bring} and \texttt{shift} to \texttt{move}, \texttt{turn} and \texttt{spin} to \texttt{rotate}, and \texttt{enlarge} and \texttt{shrink} to \texttt{resize}. Object aliases map common terms such as \texttt{box} to \texttt{cube} and \texttt{ball} to \texttt{sphere}. Table~\ref{tab:nlp_mapping} illustrates the natural language understanding capability of the command interpretation layer. For each supported action, the table presents several representative phrasings that a user might use, ranging from direct imperative commands to conversational requests and synonym-heavy expressions. All variants within a given action category produce the same structured JSON output, which demonstrates the model's ability to normalize diverse natural language into a canonical command representation.
 
\begin{table}[htp]
\centering
\caption{Examples of natural language inputs and their corresponding structured JSON outputs}
\label{tab:nlp_mapping}
\small
\renewcommand{\arraystretch}{1.25}
\begin{tabular}{p{0.08\textwidth} p{0.36\textwidth} p{0.48\textwidth}}
\hline
\textbf{Action} & \textbf{Natural Language Input} & \textbf{JSON Output} \\
\hline
 
\multirow{4}{*}{Add}
  & ``I want a traffic light here'' &
  \multirow{4}{0.48\textwidth}{\raggedright\texttt{"action": "add", "object": "traffic light", "params": "coordinates from the ray intersection"}} \\
  & ``Add a traffic light here'' & \\
  & ``Put a traffic light right there'' & \\
  & ``Place a signal light at 5, 0, 10'' & \\
\hline
 
\multirow{4}{*}{Move}
  & ``Move the cube to 3, 0, 7'' &
  \multirow{4}{0.48\textwidth}{\raggedright\texttt{"action": "move", "object": "cube", "params": \{"x":3, "y":0, "z":7\}}} \\
  & ``Shift the box over to 3, 0, 7'' & \\
  & ``Bring this to position 3, 0, 7'' & \\
  & ``Can you put that cube at 3, 0, 7?'' & \\
\hline
 
\multirow{3}{*}{Rotate}
  & ``Rotate the car 90 degrees'' &
  \multirow{3}{0.48\textwidth}{\raggedright\texttt{"action": "rotate", "object": "car", "params": \{"angle": 90\}}} \\
  & ``Turn this 90 degrees to the right'' & \\
  & ``Spin the vehicle a quarter turn'' & \\
\hline
 
\multirow{3}{*}{Resize}
  & ``Make the building twice as big'' &
  \multirow{3}{0.48\textwidth}{\raggedright\texttt{"action": "resize", "object": "building", "params": \{"scale": 2\}}} \\
  & ``Resize this by a factor of 2'' & \\
  & ``Enlarge the building to double size'' & \\
\hline
 
\multirow{3}{*}{Delete}
  & ``Delete the street lamp'' &
  \multirow{3}{0.48\textwidth}{\raggedright\texttt{"action": "delete", "object": "street lamp", "params": null}} \\
  & ``Get rid of that lamp post'' & \\
  & ``Remove this'' & \\
\hline

\multirow{3}{*}{Filter}
  & ``Filter the prefabs for traffic lights'' &
  \multirow{3}{0.48\textwidth}{\raggedright\texttt{"action": "filter", "object": "traffic lights", "params": null}} \\
  & ``Show me traffic lights'' & \\
  & ``Filter for traffic lights'' & \\
\hline
 
\end{tabular}
\end{table}

\subsection{Controller-Based Interaction}
\label{controller}
 
While voice commands provide the primary means of scene authoring in XARA, the system also supports direct manual interaction through the XR controllers. This subsection describes the ray-based selection mechanism, the manual object manipulation system, and the menu interface that together form the controller-based interaction layer.
 
\subsubsection{Ray-Casting and Object Selection}
 
Each XR controller projects a ray, a straight line extending from the controller into the virtual scene, that is used to determine which object the user is pointing at. The system calculates this ray on every rendering frame, typically 72 to 90 times per second to match the headset's refresh rate, by performing a physics query along the controller's forward direction. This physics query checks whether the ray intersects with any object in the scene, and if so, returns both the identity of that object and the exact three-dimensional point where the intersection occurs.
 
To prevent the user from accidentally selecting system objects such as their own virtual hands or interface panels, the ray is configured with a layer filter, a mechanism that restricts which categories of objects the ray can detect. Only user-created scene objects are included in the detectable category, while system infrastructure is excluded. A visible line is rendered from the controller tip to the intersection point, providing the user with continuous visual feedback showing exactly what they are pointing at. This visible line is smoothed over time using gradual interpolation between consecutive positions, a technique that blends each new position with the previous one to produce a stable beam. Without this smoothing, small natural hand tremors would cause the line to jitter visibly, making precise pointing difficult.
 
Object selection uses a toggle mechanism: pressing the designated controller button while pointing at an object selects it, and pressing the same button again deselects it. When an object is selected, the system applies a visual highlight, a colour change around the object, to indicate its active status, and the object's name appears on a display panel within the XR interface. The system maintains a persistent record of the currently selected object, which serves as a shared reference point for both manual manipulation and voice commands. For example, when a user says ``move this to 5, 0, 10,'' the word ``this'' resolves to whichever object is currently selected through the ray-based selection system.

\subsubsection{Manual Object Manipulation}
 
XARA provides a controller-based manipulation system that allows the user to move,  rotate, and scale objects directly through physical controller movements, without issuing voice commands. The system monitors the trigger buttons on both controllers and responds differently depending on which trigger, or combination of triggers, is held down.

\paragraph{Continuous Controller Manipulation}

When only the left trigger is held, the system enters movement mode. In this mode, the system records the spatial offset, defined by the distance and direction between the left controller and the selected object at the moment the trigger is first pressed. As the user moves their left hand through physical space, the object maintains this same offset and follows the controller's position in the virtual environment. This creates an intuitive grab-and-drag interaction: the object moves as though it were attached to the controller at a fixed distance, rather than snapping directly to the hand position.
 
When only the right trigger is held, the system enters rotation mode. The object rotates around its vertical axis, i.e., the up-down axis, according to changes in the direction the right controller is facing. The system records the controller's initial facing direction when the trigger is first pressed and continuously computes the angular difference between that initial direction and the controller's current direction. This angular difference is then applied to the object's orientation, so that twisting the right wrist clockwise causes the object to rotate clockwise by the same amount. The object's vertical tilt and sideways tilt are preserved from before the rotation began, ensuring that only the intended horizontal rotation is affected. As an alternative interaction method, the object can also be rotated by moving the controller laterally from side to side. This alternative control method is provided to improve usability and accommodate users for whom repeated wrist rotation may be less comfortable or intuitive. 

When both triggers are held simultaneously, the system enters scaling mode. The distance between the two controllers determines the scale of the selected object. Moving the controllers farther apart increases the object's scale proportionally, while bringing them closer together decreases it. Smoothing is applied to the resulting scale changes to reduce abrupt transitions and provide a more continuous manipulation experience.

\paragraph{Complementary Use of Voice and Gesture Interaction}

Voice and gesture interaction provide complementary approaches to object manipulation. Voice interaction is particularly useful for larger transformations that do not require fine spatial precision, such as moving an object to the cursor position. It also allows specific transformations to be expressed naturally, for example by instructing the system to rotate an object by 90 degrees or make it twice as large. Such commands allow precise transformation values to be specified directly without requiring the user to reproduce them manually through controller movement.

Gesture-based manipulation, by contrast, provides direct spatial control and is particularly suited to fine adjustment of object position, rotation, and scale. This can be useful when connecting objects such as road segments, introducing small height differences to create spatial non-uniformity, or making incremental adjustments during iterative scene development.

\paragraph{Transformation Snapping}

In addition to continuous controller manipulation, XARA provides a transformation-snapping mode. Whereas regular gesture manipulation maps controller movement continuously to object transformation, snapping maps controller movement to configurable, discrete transformation increments. This allows the relationship between physical controller movement and object transformation to be adjusted while also constraining transformations to defined increments within the world coordinate system.

Each snapped transformation has two configurable parameters: sensitivity and snap size. Sensitivity defines the amount of controller movement required to trigger one transformation increment, while snap size defines the corresponding amount by which the selected object is transformed. For example, for translation, if the sensitivity is 20 cm and the snap size is 1 m, every 20 cm of controller movement along an axis translates the selected object by 1 m along the corresponding world axis. When the user first holds down the left trigger to initiate snapped translation, the system establishes a local origin at the controller's position and calculates subsequent displacement relative to this reference position. A one-to-one relationship can also be configured by setting equivalent controller-movement and transformation increments. Unlike continuous manipulation, however, the resulting object positions remain constrained to discrete increments on the world grid.

When transformation snapping is enabled, the system provides visual guides corresponding to translation, rotation, and scaling. These guides communicate the configured snap increments and provide visual feedback during manipulation.

\paragraph{Snapped Translation}

When snapped translation is initiated, a grid is displayed at the selected object's origin and colour coded by axis. The grid's units are the same size in the world as the snap size. The grid provides a visual representation of the available translation increments and helps the user assess how the configured snap size corresponds to the surrounding scene. It is rendered above other scene geometry so that it remains visible during manipulation. To translate the object, the user moves the controller in the desired direction, with each completed sensitivity increment producing the corresponding snapped movement of the selected object. 

\paragraph{Snapped Rotation}
When snapped rotation is initiated, a transparent circular guide is displayed with a marker at each multiple of the configured snap size. For example, a snap size of 90° produces markers at 0°, 90°, 180°, and 270°. For rotation, the snap size is expressed in degrees, while sensitivity is expressed as a linear controller displacement. Unlike continuous rotation, snapped rotation does not require the user to reproduce the desired angular change through wrist rotation. Instead, lateral movement of the controller is mapped to discrete rotational increments. When the user initiates snapped rotation, the system stores both the controller's initial position and its right vector. On each frame, the displacement vector between the initial and current controller positions is projected onto the stored right vector. The resulting lateral displacement determines the number of rotation increments applied to the selected object. For example, if the sensitivity is 10cm and the snap is 10°, then if moving the controller 25cm to the right in the positive direction results in an object rotation of +20°. In addition to the markers on the circular guide, an inner blue indicator shows the amount of rotation added or subtracted since the transformation began. This provides a visual reference to the object's initial orientation and allows subsequent rotation to be assessed relative to that starting point. Unlike the translation grid, the rotation guide is rendered as regular scene geometry rather than as a persistent foreground overlay, as rendering it above all other objects was found to be visually disruptive.

\paragraph{Snapped Scaling}

\begin{figure}[htp]
    \centering

    \begin{subfigure}[b]{0.31\textwidth}
        \centering
        \includegraphics[width=\textwidth]{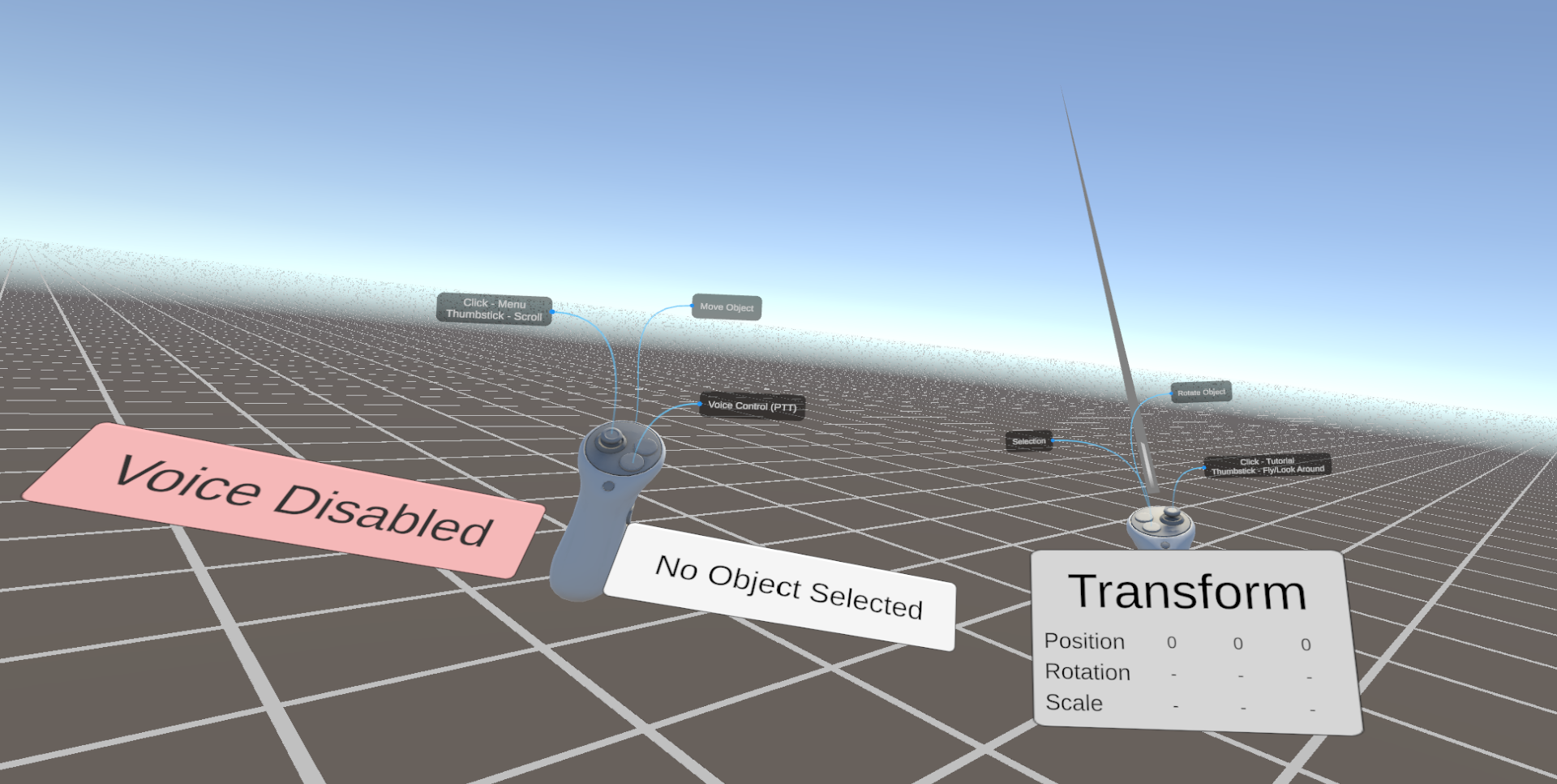}
        \caption{Transformation menu with explanation of each controller option.}
        \label{fig:transform5}
    \end{subfigure}
    \hfill
    \begin{subfigure}[b]{0.31\textwidth}
        \centering
        \includegraphics[width=\textwidth]{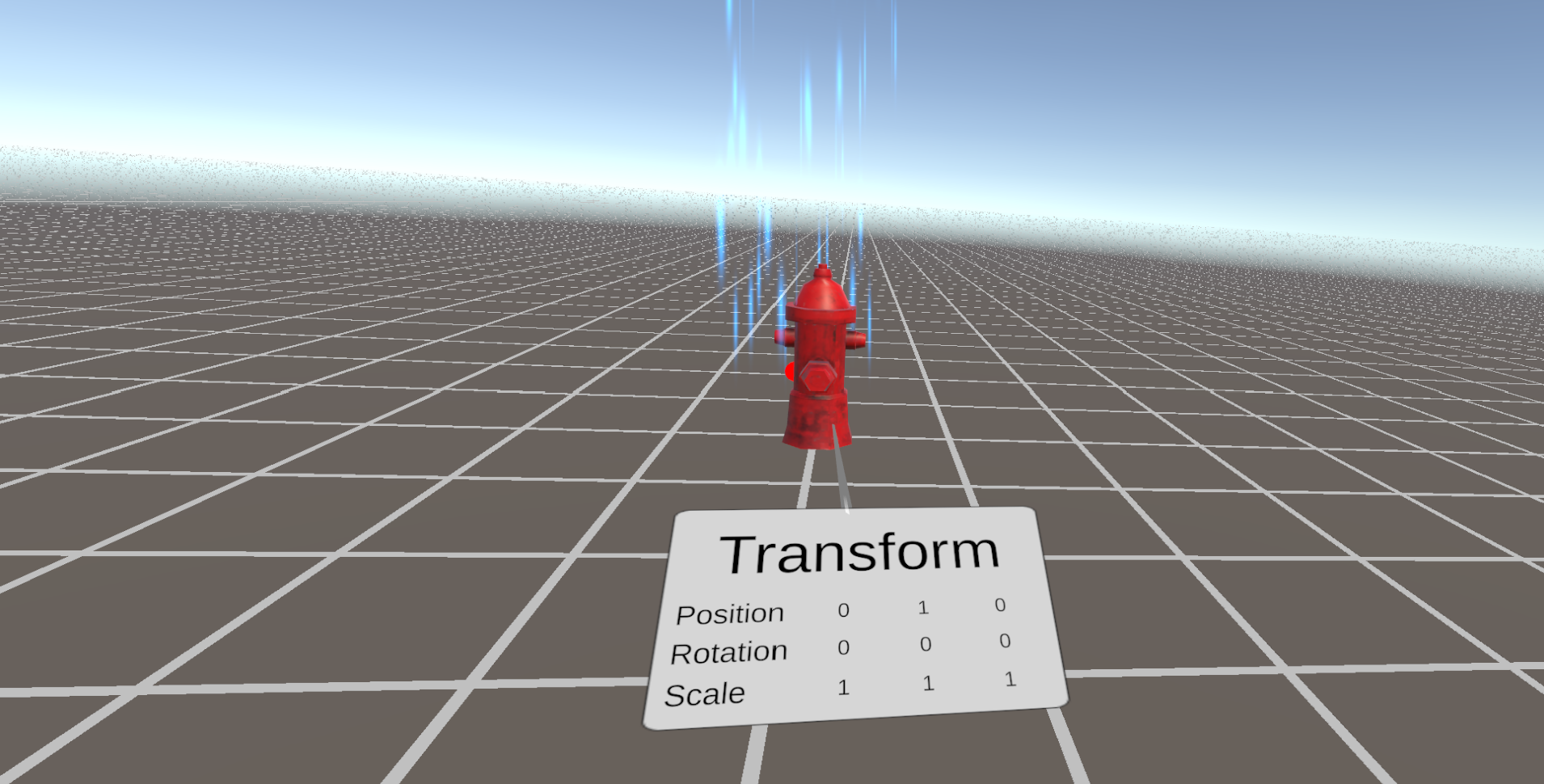}
        \caption{Snapped transformation based on values.}
        \label{fig:transform4}
    \end{subfigure}
    \hfill
    \begin{subfigure}[b]{0.31\textwidth}
        \centering
        \includegraphics[width=\textwidth]{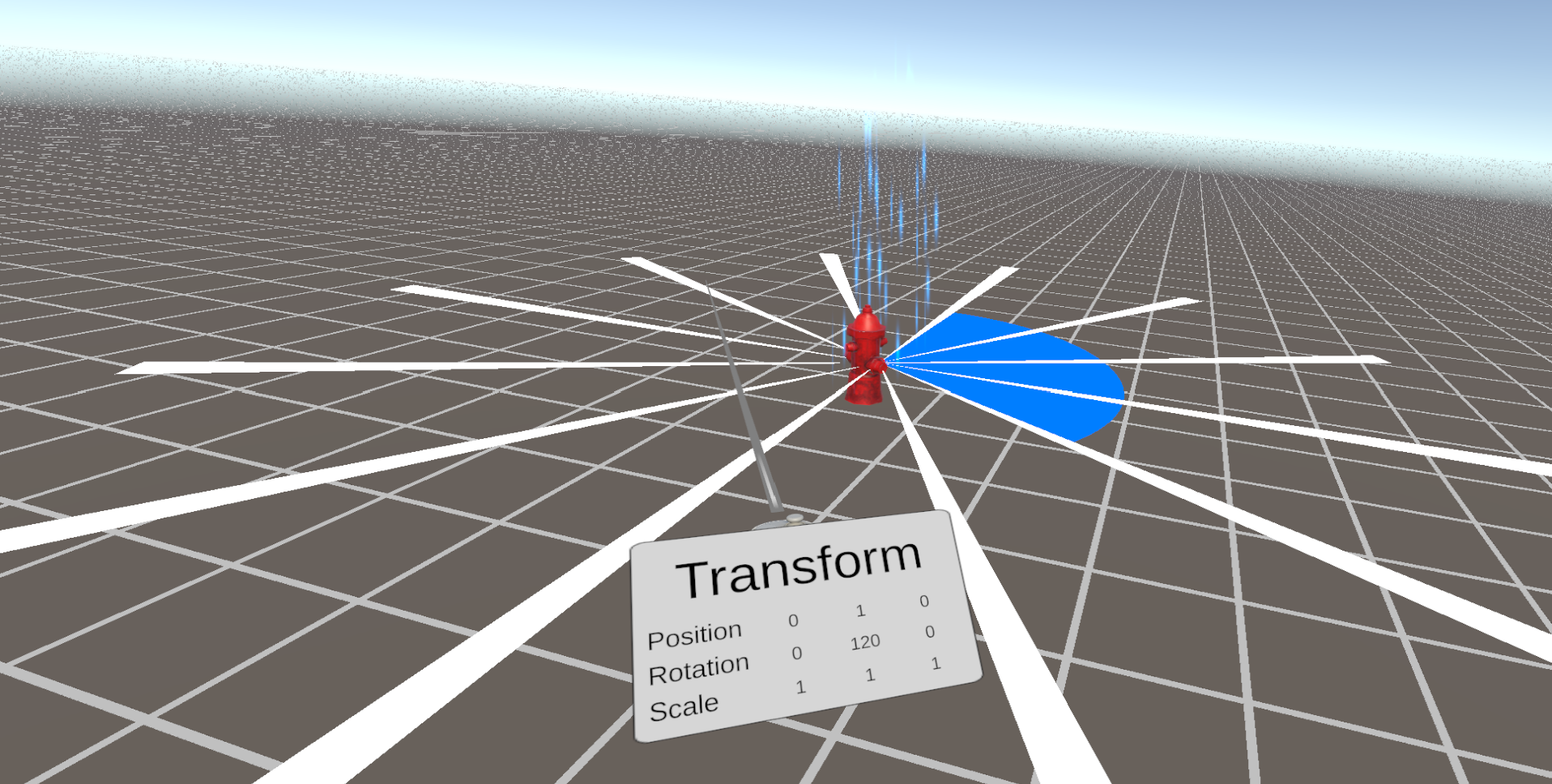}
        \caption{Snapped transformation with rotation option.}
        \label{fig:transform1}
    \end{subfigure}

    \vspace{0.5em}

    \begin{subfigure}[b]{0.31\textwidth}
        \centering
        \includegraphics[width=\textwidth]{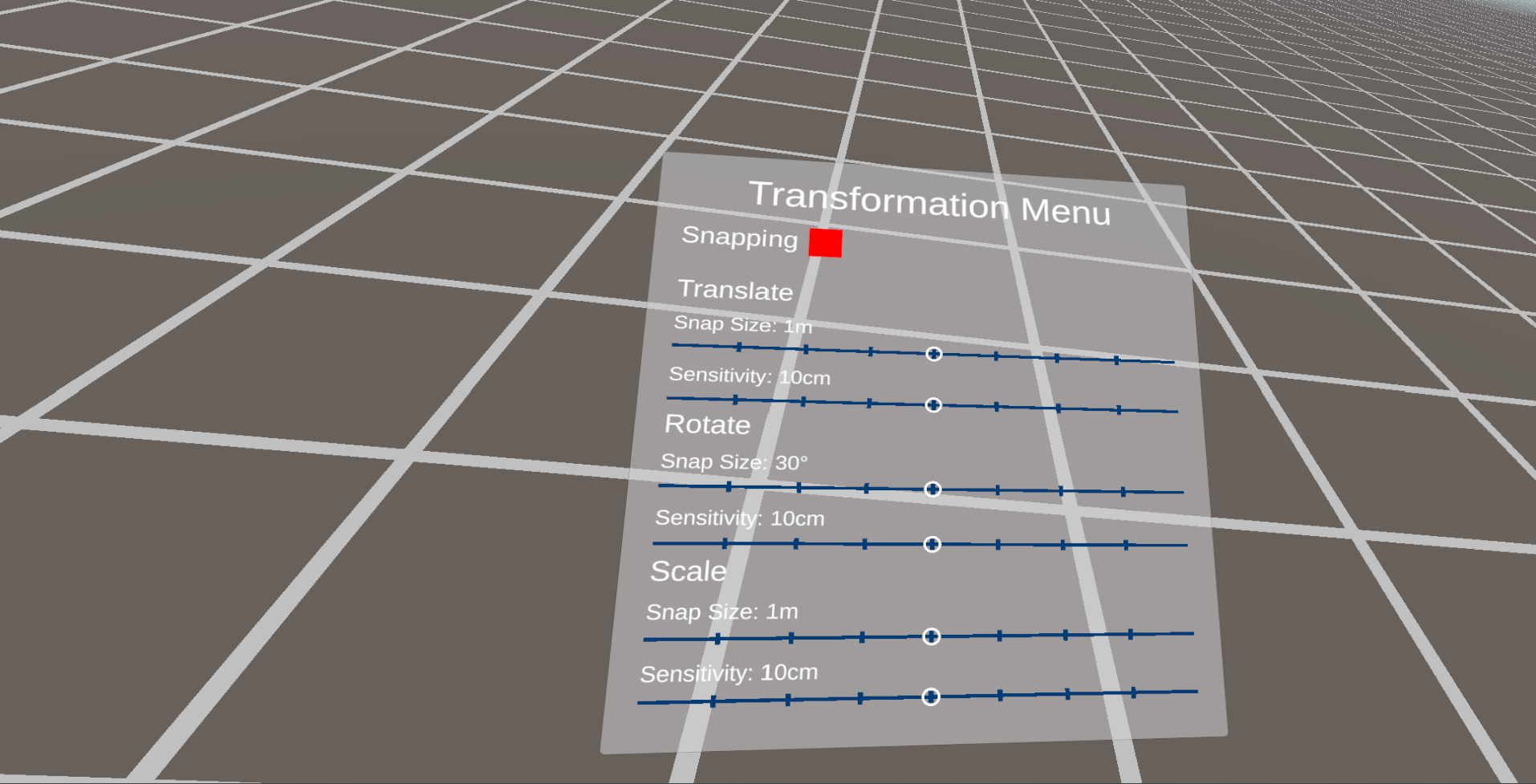}
        \caption{Snapped transformation menu with sliders.}
        \label{fig:transform2}
    \end{subfigure}
    \hfill
    \begin{subfigure}[b]{0.31\textwidth}
        \centering
        \includegraphics[width=\textwidth]{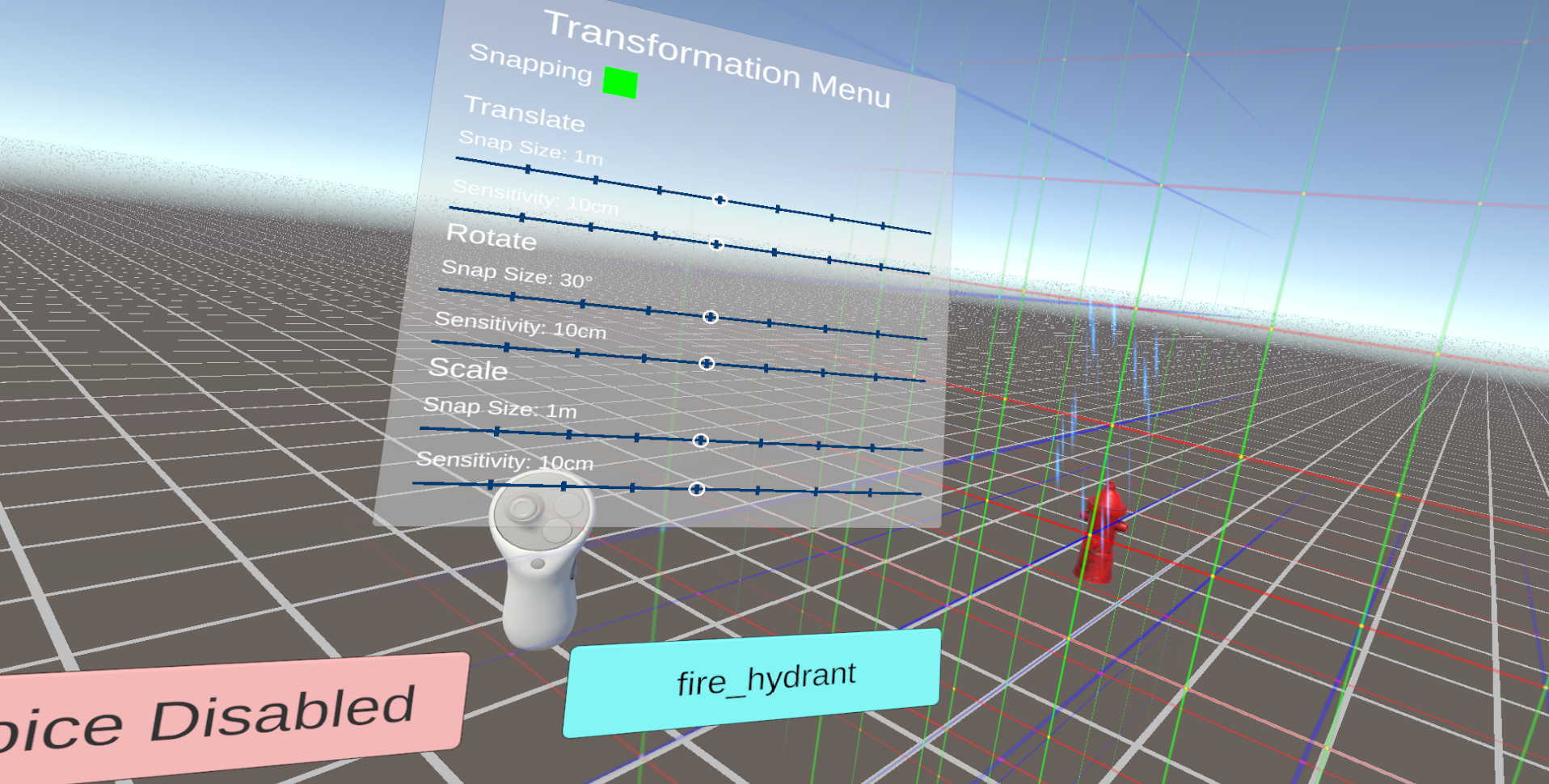}
        \caption{Snapped transformation visualization.}
        \label{fig:transform3}
    \end{subfigure}
    \hfill
    \begin{subfigure}[b]{0.31\textwidth}
        \centering
        \includegraphics[width=\textwidth]{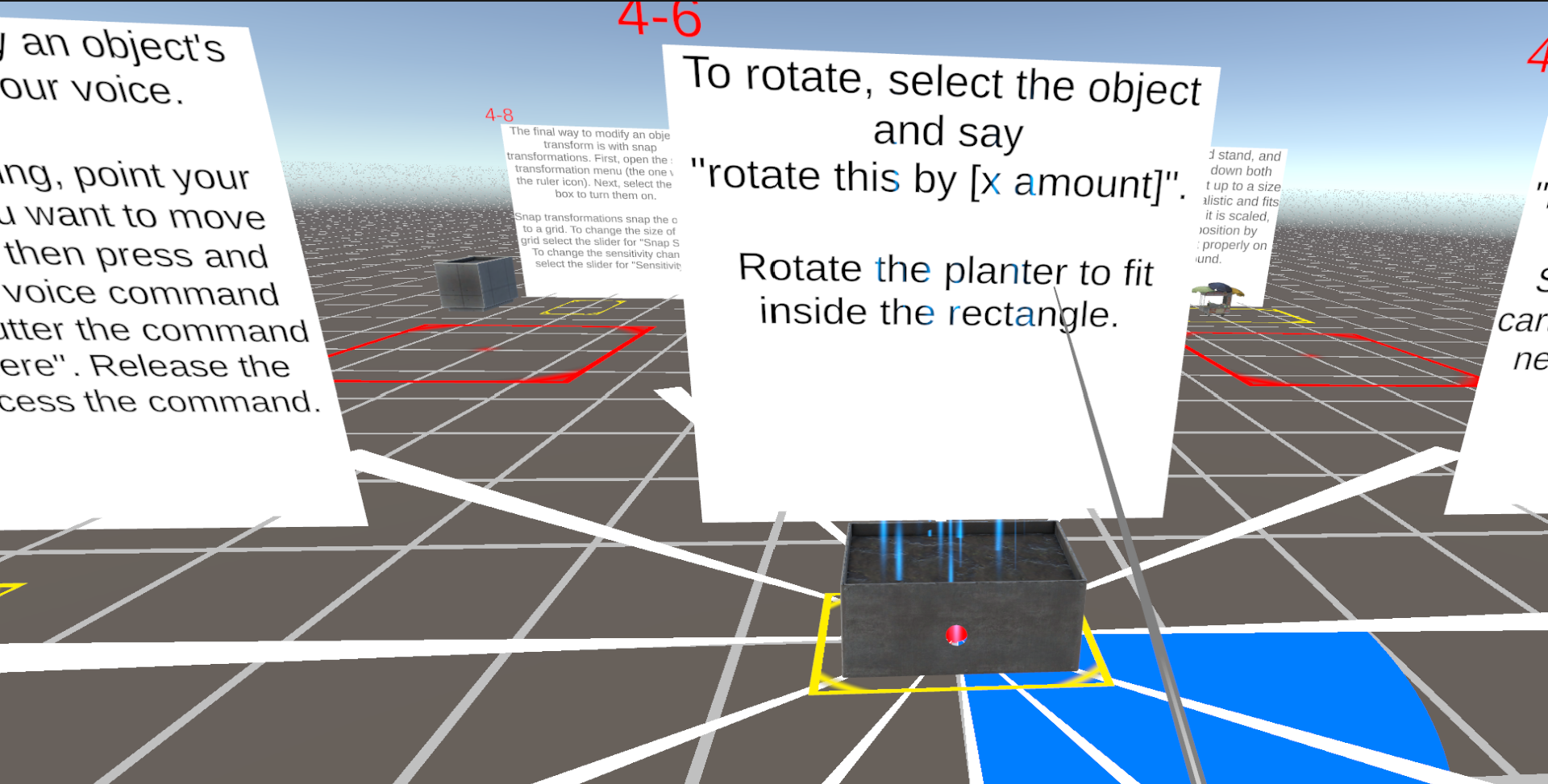}
        \caption{Continuous controller-based rotation.}
        \label{fig:rotate}
    \end{subfigure}

    \caption{Controller-based object manipulation and visual feedback provided by XARA during continuous rotation and snapped translation, rotation, and scaling.}
    \label{fig:object_transformations}
\end{figure}

Snapped scaling extends the same sensitivity-and-increment approach used for translation and rotation to the scaling operation. Whereas continuous two-handed scaling is controlled by the distance between the two controllers, snapped scaling uses controller displacement relative to an initial reference position to apply discrete changes in scale. When the user holds down both triggers and initiates snapped scaling, the system records the position of the right controller. The displacement of the right controller from this initial position is then evaluated against the configured sensitivity. For each sensitivity increment crossed, the configured snap size is added to or subtracted from the selected object's scale. To provide visual feedback during this interaction, a ruler-like guide is displayed along the relevant movement axis. A notch is shown at each sensitivity increment, allowing the user to see how many scaling increments have been applied. As with the rotation guide, this visualization is rendered as regular scene geometry rather than as a foreground overlay.
The visual feedback mechanisms associated with snapped translation, rotation, and scaling are illustrated in Figure~\ref{fig:object_transformations}.

\subsubsection{Main Menu}
 \label{mainmenu}
The XR menu system provides access to functions that are not suited to voice input, such as creating a new scene, loading a previously saved scene, and saving the current scene state. The menu is displayed as a flat panel, a two-dimensional interface surface, floating at a fixed position relative to the left controller. The user interacts with the menu using the controller's ray in the same way one would use a laser pointer on a projected screen, pointing the ray at a button highlights it visually, and pressing the controller trigger activates it.
 
Navigation between menu panels is performed using a selection wheel controlled by the left thumbstick. Pressing down on the thumbstick opens the selection wheel, which is divided into segments corresponding to the available menus. Each segment contains an icon representing its associated menu, including the main scene-management menu, the scene-loading menu, and the prefab menu. The user selects a menu by moving the thumbstick towards the corresponding segment. The selected segment is highlighted in blue, and the associated menu panel is then displayed to the user. This approach allows direct switching between menus while minimising the number of controller inputs required for interface navigation. 

\begin{figure}[htp]
    \centering

    \begin{subfigure}[b]{0.32\textwidth}
        \centering
        \includegraphics[width=\textwidth]{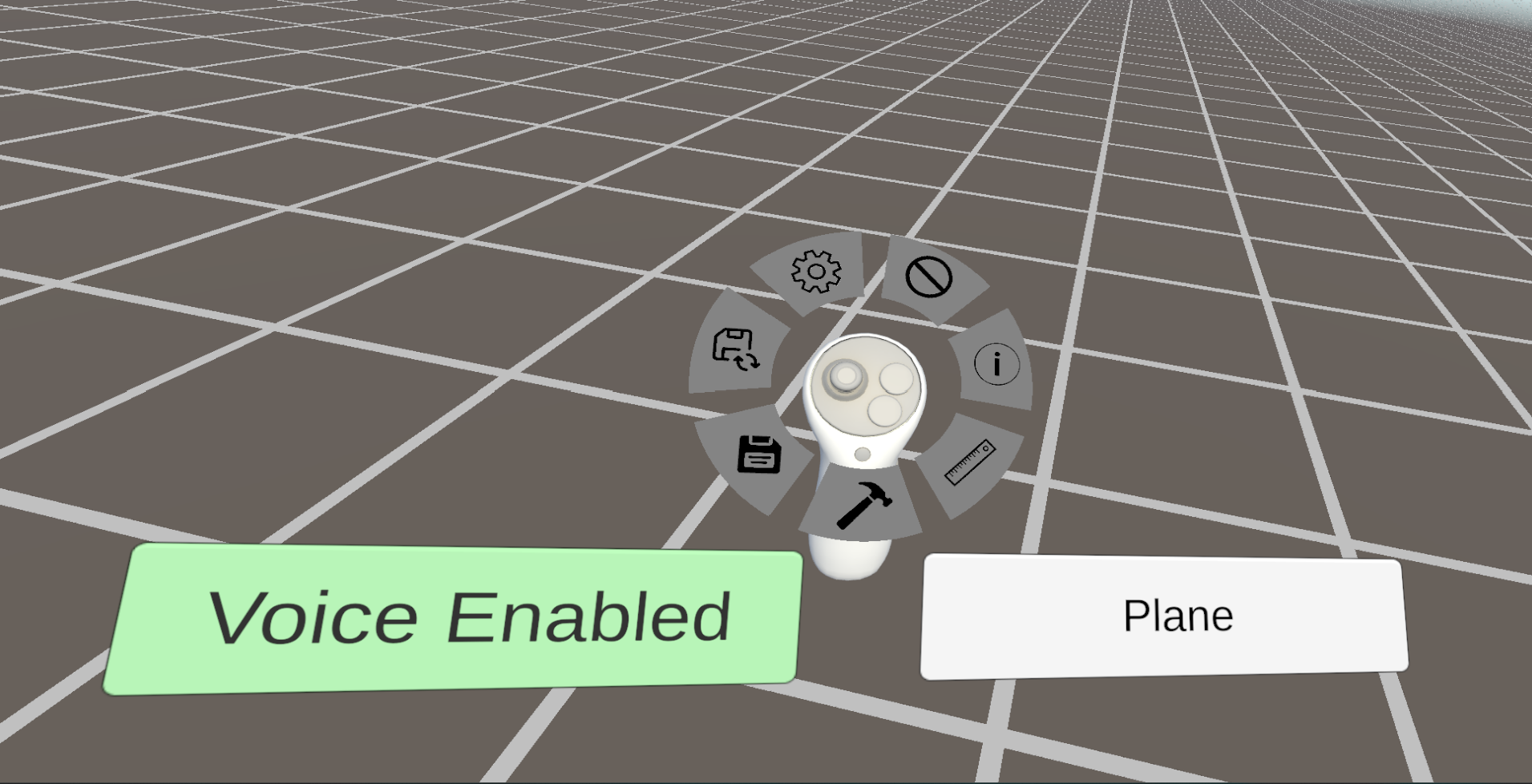}
        \caption{Radial selection wheel used to switch between menu panels.}
        \label{fig:menu_radial}
    \end{subfigure}
    \hfill
    \begin{subfigure}[b]{0.32\textwidth}
        \centering
        \includegraphics[width=\textwidth]{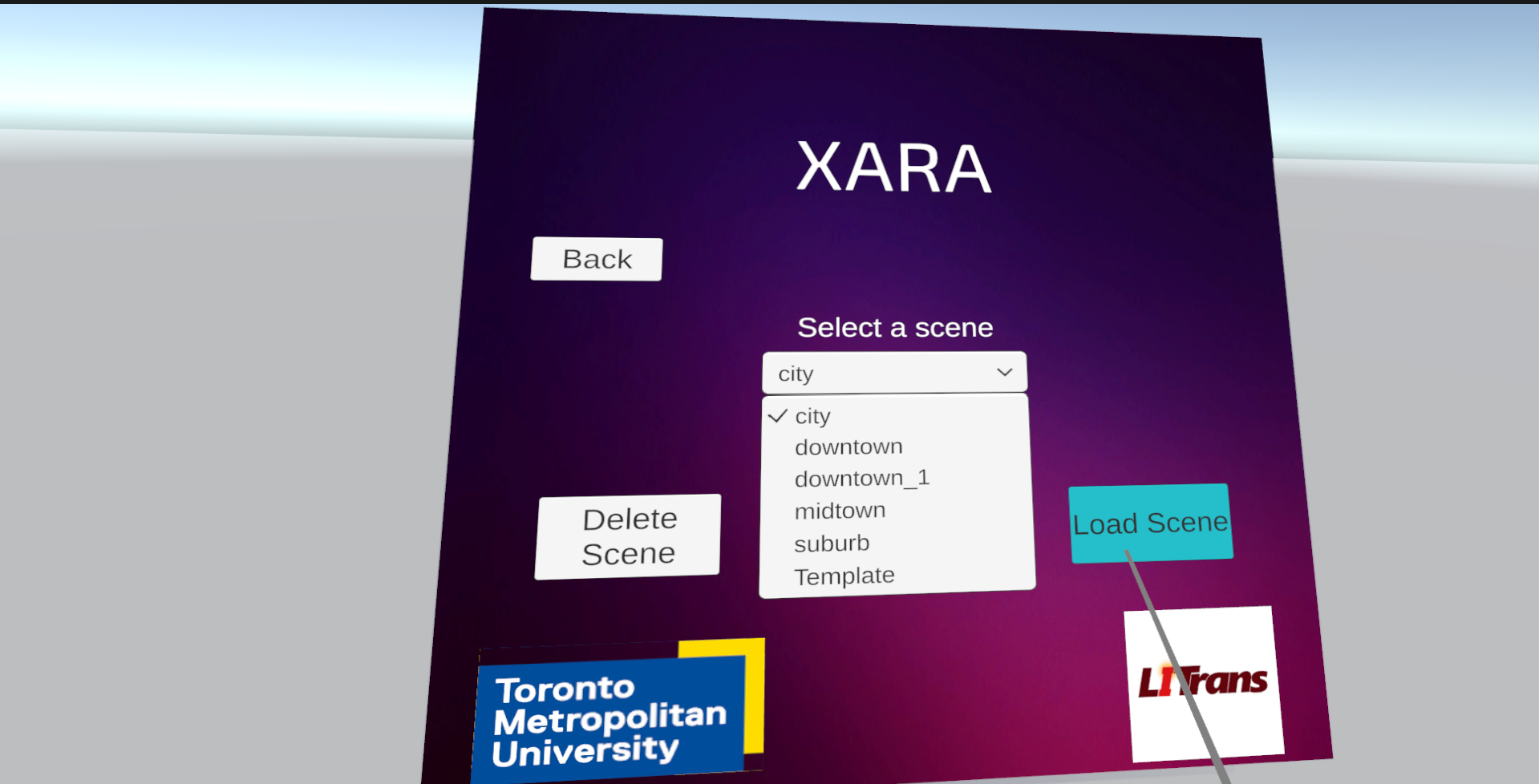}
        \caption{Main menu interface within the XR environment.}
        \label{fig:main_menu_view}
    \end{subfigure}
    \hfill
    \begin{subfigure}[b]{0.32\textwidth}
        \centering
        \includegraphics[width=\textwidth]{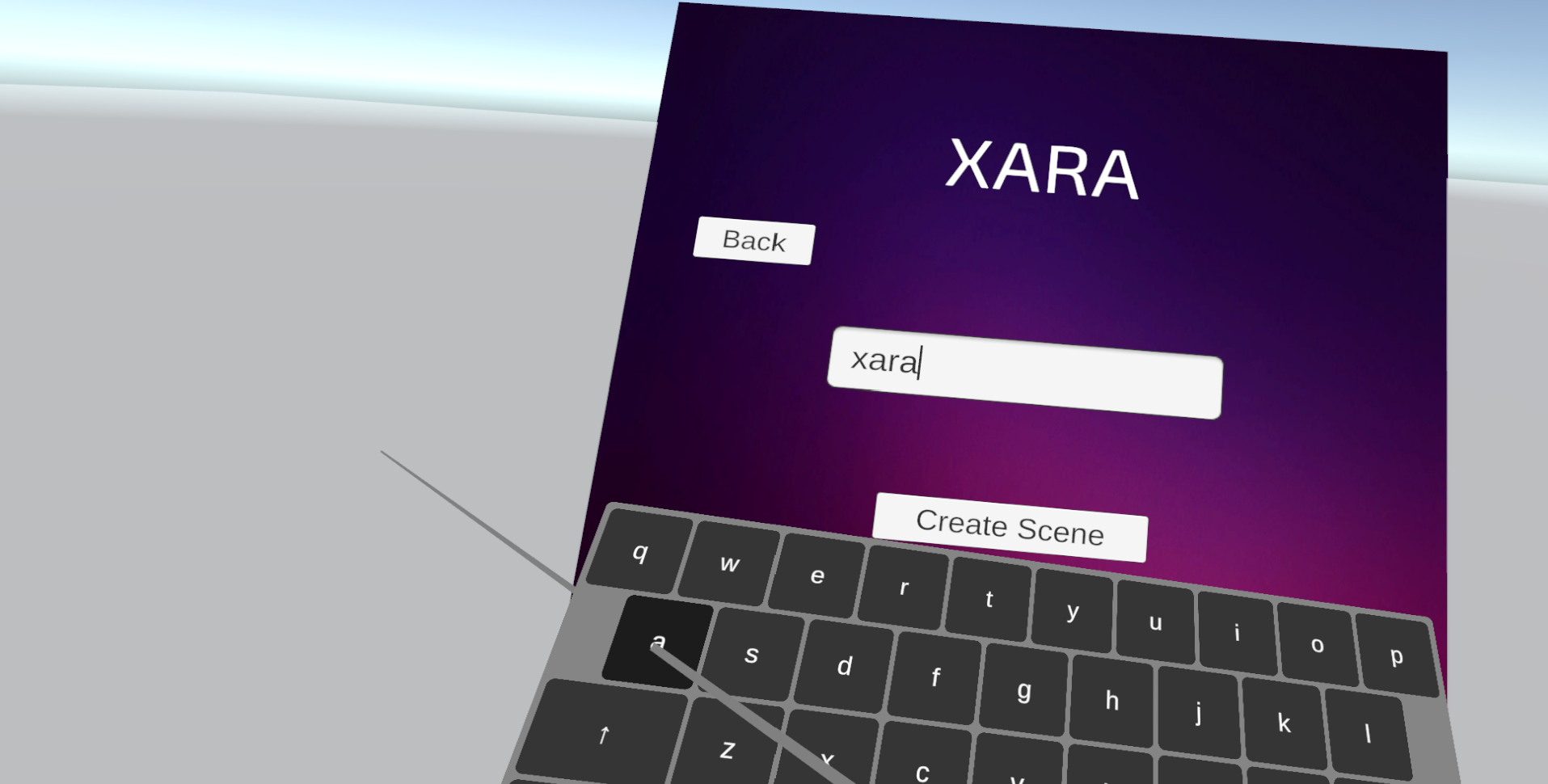}
        \caption{Scene-management menu used for creating new scenes.}
        \label{fig:scene_menu_view}
    \end{subfigure}

    \caption{XR menu system in XARA. The radial selection wheel enables direct switching between menu panels, while the main menu and scene-management menu provide access to scene creation and loading functions.}
    \label{fig:main_menu}
\end{figure}

The menu navigation mechanism and the main scene-management interfaces are illustrated in Figure \ref{fig:main_menu}.

\subsubsection{Prefab Menu}
\label{prefabmenu}
 
\begin{figure}[htp]
    \centering
        \includegraphics[width=0.5\textwidth]{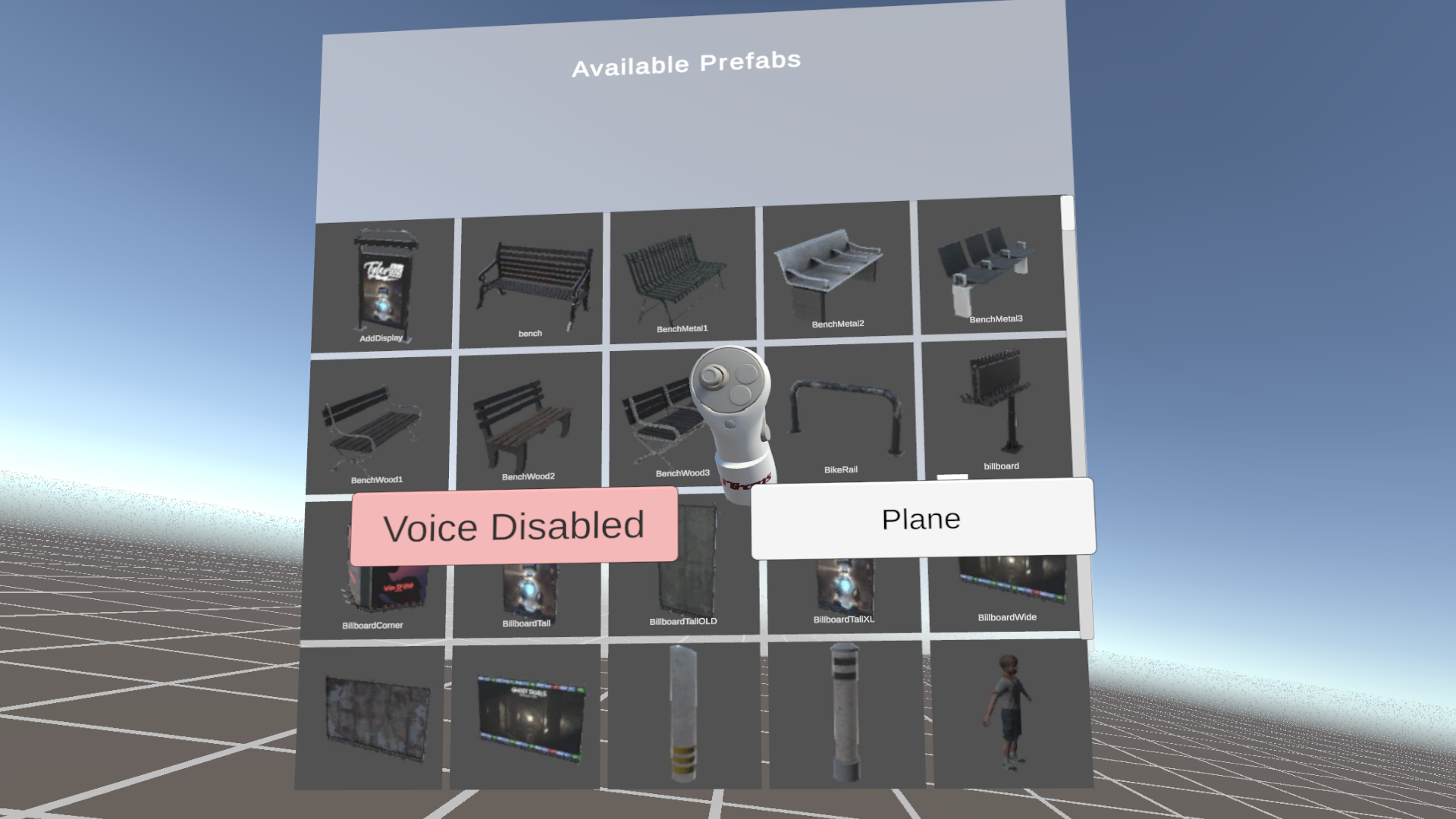}
    \caption{Prefab menu for XARA.}
    \label{fig:mfab_menu}
\end{figure}

The prefab menu (see Figure \ref{fig:mfab_menu}) provides a visual catalogue of the objects available for placement within the virtual scene, with each prefab represented by a thumbnail image. The user selects a prefab by directing the controller ray towards its thumbnail and pressing the trigger, which registers the corresponding prefab as the active object for placement. The user can then direct the controller ray towards the desired location in the virtual environment and press the trigger again to instantiate the selected prefab The placement position is determined by the ray's intersection with existing scene geometry, such as the ground plane or the surface of another object, allowing for precise spatial positioning. 

Using voice commands there is an option to filter through the prefab menu to only display prefabs that are relevant for your use. The command syntax is very loose, being simply "Filter [filter]", with [filter] being any natural language filter. When the voice command is processed a second call is made to the reasoning model which passes the filter and a string containing every object name formatted as ``object1,object2,object3,...,objectX''. The reasoning model is prompted to return a string of the format "obj1,obj2,...,objN" which contains all prefabs relevant to the filter. For example, using the command "Filter for buildings", the reasoning model may return a string of the form "building1,skyscraper,residential\_building". This string of object names is then processed and the prefab menu is cleared and repopulated with only the objects returned. To clear the filter the command "Clear filter" is used which simply clears all the currently visible prefabs and re-populates the menu with all of the objects. A reasoning model was used instead of a hard classification objects as it is more natural to use and in theory should support a limitless number of filters. 

\subsection{Scene Generation and Manipulation}
\label{scene}
 
Once the language processing pipeline produces a structured JSON command, or a controller-based action triggers a scene modification, the scene generation subsystem translates that instruction into a visible change within the virtual environment. This subsystem operates within the Unity game engine and is built around a central command dispatcher, a software component that receives incoming commands, determines the requested action type, and delegates execution to the appropriate procedure. Because the language processing pipeline runs on a separate execution thread from the rendering engine, incoming commands cannot be applied to the scene directly; instead, they are placed into a thread-safe queue, a data structure designed to allow one thread to add items while another thread removes them without either thread interfering with the other. During each rendering frame, the main Unity thread retrieves the next command from this queue and executes it, ensuring that all scene modifications occur in a controlled, sequential manner that does not conflict with the rendering cycle.
 
The virtual environment contains a library of pre-built three-dimensional models, referred to in Unity terminology as prefabs. These prefabs represent objects commonly used in transportation scenarios, including vehicles, traffic signs, pedestrians, buildings, road segments, and street furniture. At application startup, the system performs a recursive scan of the project's asset directory, traversing all subdirectories, and constructs an indexed catalogue that maps each object's name to its storage location within the project. This catalogue is stored as a dictionary data structure, which allows the system to look up any object by name in constant time regardless of how many objects are available.
 
When the user issues a voice command such as ``add a car,'' the resulting JSON command triggers a lookup against this catalogue. If an exact name match is not found, the system attempts several normalized variants, such as replacing spaces and hyphens with underscores, before conducting a broader search across all resource subdirectories. Once the matching prefab is identified, the system creates a new instance, an independent copy, of that model and places it in the scene. Placement follows one of two modes: if the command includes explicit three-dimensional coordinates, for example ``add a car at 5, 0, 10,'' the object is positioned at those world-space coordinates where the three numbers represent the left-right, vertical, and forward-backward positions respectively; alternatively, if the user says ``here,'' the system casts a ray from the XR controller into the scene and places the object at the point where that ray intersects with a surface, such as the ground plane. 

Before the system can manipulate an object, whether by voice or gestures, it must first determine which object the user is referring to. XARA supports multiple methods for this target resolution. The user can refer to an object by its assigned name, for example ``move the car,'' in which case the system searches through all objects currently present in the scene hierarchy to find one whose name matches the spoken identifier. Alternatively, the user can say ``this'' to refer to whichever object is currently highlighted by the controller's ray-based selection system, as described in Section~\ref{controller}. This dual addressing approach allows users to interact with the scene using whichever reference method is most convenient, naming is useful when the target object is far away or visually occluded, while selection-based reference is useful when pointing is faster than recalling an object's name.
 
Once the target object has been identified, the system applies the requested transformation. Movement commands update the object's position in three-dimensional space, setting it either to the coordinates specified in the command or to the current intersection point of the controller's ray with the scene. Rotation commands turn the object around its vertical axis by a specified number of degrees: for example, ``rotate the car 90 degrees'' applies a 90-degree rotation that would turn a vehicle from facing forward to facing sideways. Internally, this rotation is represented using quaternions, a mathematical representation of orientation that avoids computational problems, such as gimbal lock, that can occur with simpler angle-based representations. Scaling commands multiply the object's size uniformly along all three spatial axes by the specified factor; a factor of 2 doubles the object's dimensions, while a factor of 0.5 halves them. All voice-driven transformations are applied instantaneously in a single operation, in contrast to the continuous frame-by-frame updates produced by manual controller manipulation.
 
To support the iterative nature of experiment design, the system maintains a record of each object's spatial state, comprising its position, rotation, and scale, prior to every transformation. These records are stored in per-object history stacks, a data structure in which each new entry is placed on top of the previous one, preserving the chronological sequence of past states. When the user invokes the undo function through the XR menu, the system retrieves the most recent entry from the stack and restores the object to that previous state. Multiple consecutive undo operations step progressively further back through the history. A corresponding redo function reverses an undo by restoring the state that was most recently reverted. A short cooldown period is enforced between consecutive undo and redo operations to prevent rapid repeated activations from a single prolonged button press. Deletion operations maintain a separate history mechanism that allows deleted objects to be restored.
 
Experiment scenarios can be saved to disk and reloaded in subsequent sessions. When the user triggers a save operation through a settings menu in the scene, the system iterates through all user-created objects in the scene and records each object's name, type tag, position, rotation, and scale into a structured JSON file. System-level objects, including the XR controllers, interface menus, lighting rigs, and camera objects, are identified by their tags and excluded from the export to ensure that the saved file contains only the experiment-relevant scene content. When the application is launched, the scene loader reads the most recent JSON file and reconstructs the saved environment by creating new instances of each recorded object and setting their spatial properties to the saved values. This persistence mechanism serves three purposes as it enables users to build complex scenarios incrementally over multiple work sessions rather than completing them in a single sitting; it allows experiment configurations to be shared with collaborators by transferring the JSON file; and it ensures that experimental conditions are precisely reproducible, which is essential for running repeated trials with consistent scene parameters.

\subsection{User Experience Features}
\label{ux}
Several UX features are added to XARA to ensure seamless navigation and ease of use. An undeletable attribute is added to make sure that users are not able to accidentally delete system components, in addition to the base plane which is useful for users placing objects into the world. Initially, object transforms were only visible when the transform information menu was open, but in testing we found that this information was useful to have at all times. To make it more accessible a transform display panel was added on the right hand. When an object is selected, it will display the position, rotation, and scale of that object, otherwise it will show the position the right hand ray is pointing at - which is especially useful for translation using voice commands. 

Initially, changing menus was done by moving backwards and forwards through a cyclic list of menus until arriving at the one you wanted. While this worked, it felt somewhat slow and clunky to go through every menu until arriving at the one you required. This also tied up two inputs on the left hand - pushing the stick to the left and right. Swapping between menus was redesigned to use a radial menu instead which allows users to directly select the menu they want. When the radial menu opens it changes the input mapping context so that the left stick is used only for menu selection. When it is closed the left stick is freed up and the entire vector 2 input is available. This leaves room in the future for a feature that makes use of the full input instead of only being able to receive two booleans - stick up and stick down.

Lastly, in testing we found that users had a hard time remembering all the controls in XARA. To make it easier to see what each input does, hovering labels were added that have a spline curve attached to them which point directly to the input that action is mapped to (see Figure \ref{fig:transform5}). For example, the right stick has a label which states it should be pushed forwards to move and pushed to the left and right to rotate the camera. 

\subsection{Orientation Session}
\label{orientation}

To facilitate familiarization with XARA, a dedicated orientation scene was developed. The scene is divided into five stages covering movement, controller operation, object creation, object manipulation, and editing operations. The orientation session is important for ensuring that participants begin the experiment with a consistent baseline understanding of the system, thereby reducing variability caused by unfamiliarity with the interface or control mechanisms and allowing subsequent behaviour to more accurately reflect the experimental tasks themselves. Each stage introduces a specific group of XARA functions through a combination of explicit instruction and implicit reinforcement. New interactions are first introduced directly, while subsequent tasks require users to reuse previously learned interactions without repeating the complete instruction. This progressive structure allows users to become familiar with the system while completing increasingly complex tasks. A complete walkthrough of the orientation session and the interactions described in this section is available in the accompanying XARA tutorial video.\footnote{XARA Tutorial: \url{https://youtu.be/A6PlUJB6QJU}}

The first stage introduces movement in XARA and consists of two stations. At the first station, the user is instructed to fly forward to the second station, providing an explicit introduction to the flight control. At the second station, the user is shown how to rotate the viewing direction and is required to use this rotation to locate the following station. Reaching that station also requires the previously introduced flight control, thereby reinforcing movement while introducing camera rotation.

The second stage consists of a single station that introduces the controller layout. Pressing the right thumbstick displays labels connected by splines to the corresponding controller inputs, allowing the user to identify the controls and their associated functions directly within the XR environment.

The third stage consists of three stations and introduces object creation through the prefab menu. At the first station, users are instructed to open the prefab menu and place a fire hydrant, one of the first objects displayed in the catalogue. This provides a direct introduction to the prefab-selection and placement workflow. At the second station, users scroll through the menu to locate the prefab named ``road\_basic'' and place it within a predefined outline. In addition to introducing menu navigation, this task reinforces the use of flight to reposition the user when interacting with larger scene elements. It also prepares users for the subsequent experimental task, in which road elements are used to construct a basic downtown environment.

At the final object-creation station, users are introduced to voice interaction through prefab filtering. They are instructed to use the filter command to identify a traffic-light prefab, place it within an outline, and then use the clear-filter command to restore the full prefab catalogue. In addition to introducing voice-based filtering, this task familiarizes users with the general structure of XARA voice commands, following the pattern command $\rightarrow$ (target) $\rightarrow$ descriptor. The relatively simple filter interaction therefore serves as an accessible introduction to natural-language interaction with the system.

The fourth stage focuses on object manipulation and consists of ten stations, making it the most extensive component of the orientation session. The first station introduces object transforms and the transform display. Users begin by directing the controller ray toward the centre of a target and reading the corresponding transform information. They then select three copies of a fire hydrant, each configured with different transform values, to observe how changes in position, rotation, and scale affect an object within the virtual environment. This introduces object selection explicitly while also establishing the transform display on the right controller as a reference that can be used throughout subsequent manipulation tasks.

The remaining stations introduce translation, rotation, and scaling using the three manipulation methods available in XARA: continuous controller gestures, voice commands, and snapped transformations. The same underlying transformation tasks are repeated using each interaction method so that users can focus on learning a new control mechanism while working toward a spatial outcome with which they are already familiar. Continuous gesture manipulation is introduced first, followed by voice commands and then snapped transformations. This progression provides repeated practice with the core manipulation operations while gradually introducing more specialized interaction techniques.

For translation tasks, users move a planter from its initial position into a predefined outline. When completing the task using voice interaction, they are instructed to use the command ``move this here'', reinforcing the command $\rightarrow$ (target) $\rightarrow$ descriptor structure introduced earlier. For rotation tasks, users rotate a planter until it aligns with the orientation represented by an outline. For scaling tasks, users resize an object to fit within an outline and then translate it vertically until it is positioned correctly at ground level. This final step implicitly reviews translation and demonstrates that several transformations can be applied sequentially to obtain the desired scene configuration.

The final stage introduces editing and history operations and consists of three stations. Users are first taught to delete an object through the interface after selecting it, reinforcing the requirement to establish an active object selection before performing object-specific operations. They are subsequently shown how to perform deletion using a voice command, providing further practice with the voice-command structure. In the final task, users transform a fire hydrant to fit within an outline and then delete it. They are then instructed to undo both operations and subsequently redo the transformation without reapplying the deletion. This task introduces the undo and redo functions and demonstrates how a sequence of scene modifications can be selectively reversed and reapplied during scene development.

\subsection{Asset Generation and Manipulation}
\label{assets}
To support on-demand 3D asset creation, XARA will implement TRELLIS, an open-source generative AI model, developed by Microsoft Research, to create unique and customized 3D assets for scenes, which uses text prompts to generate assets \citep{xiang2024trellis}. TRELLIS uses a unified structured latent 3D representation (SLAT) that can be decoded into various output formats such as 3D meshes, Gaussian splats, and radiance fields \citep{xiang2024trellis}.

TRELLIS is currently set up as part of the XARA system and its integration is actively being implemented. The planned integration connects TRELLIS to the existing voice command subsystem of XARA's, allowing users to generate assets dynamically during a VR session. The asset generation feature reuses the voice command subsystem described in Sections 3.2 and 3.3. Unlike standard scene manipulation commands, however, asset generation prompts are not constrained by the action/object/params schema. Instead, the LLM reformats the transcript into a free-form text prompt, which is sent as a JSON payload in a POST request to a FastAPI server.
An example prompt is shown below and the related visualization in Figure \ref{fig:assetgen}:

\begin{verbatim}
{
  "prompt": "a black electric sedan."
}
\end{verbatim}

\begin{figure}[htp]
    \centering

    \begin{subfigure}[b]{0.45\textwidth}
        \centering
        \includegraphics[width=\textwidth]{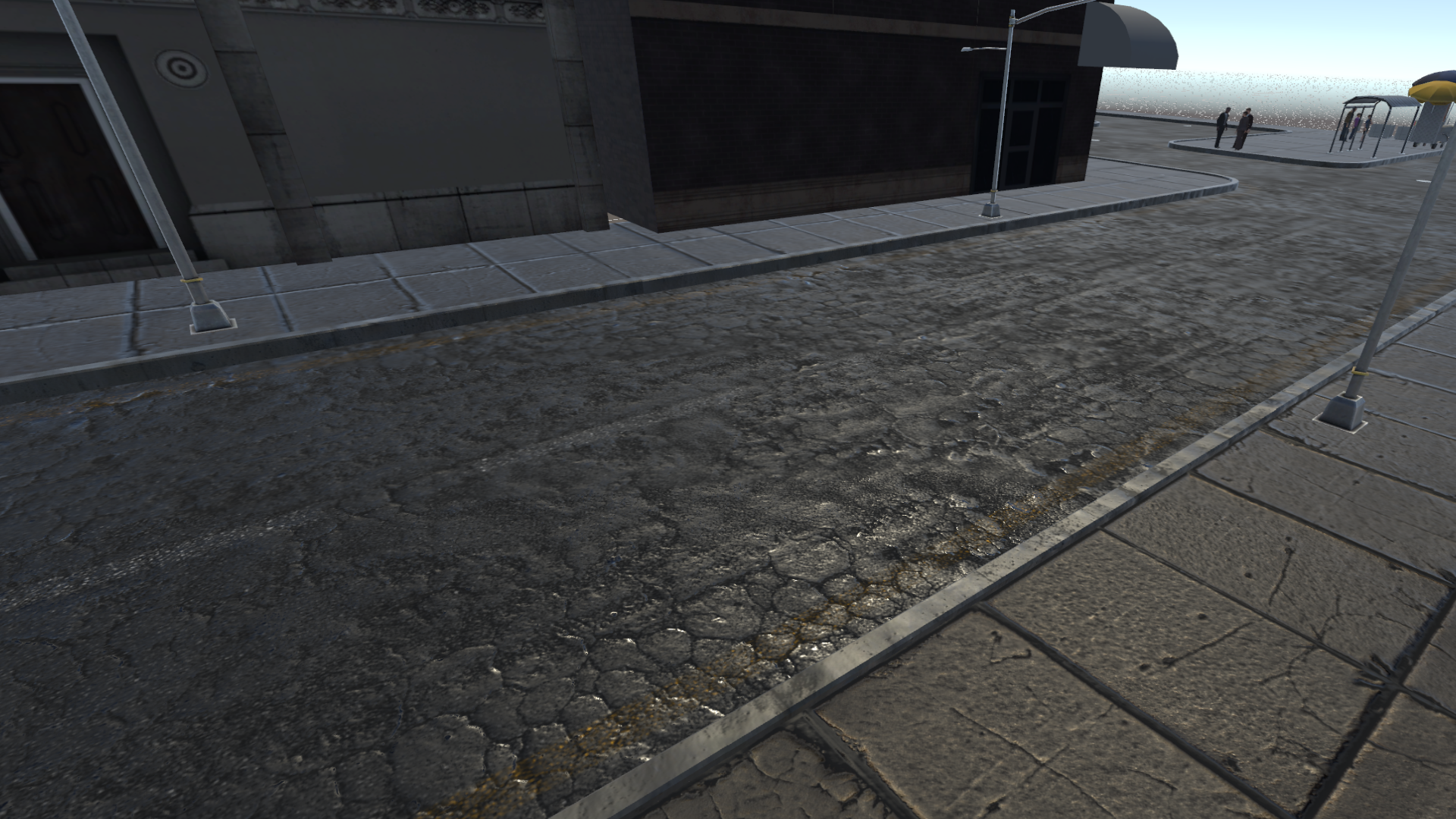}
        \caption{Scene before the prompt for asset generation.}
        \label{fig:assetgen1}
    \end{subfigure}
    \hfill
    \begin{subfigure}[b]{0.45\textwidth}
        \centering
        \includegraphics[width=\textwidth]{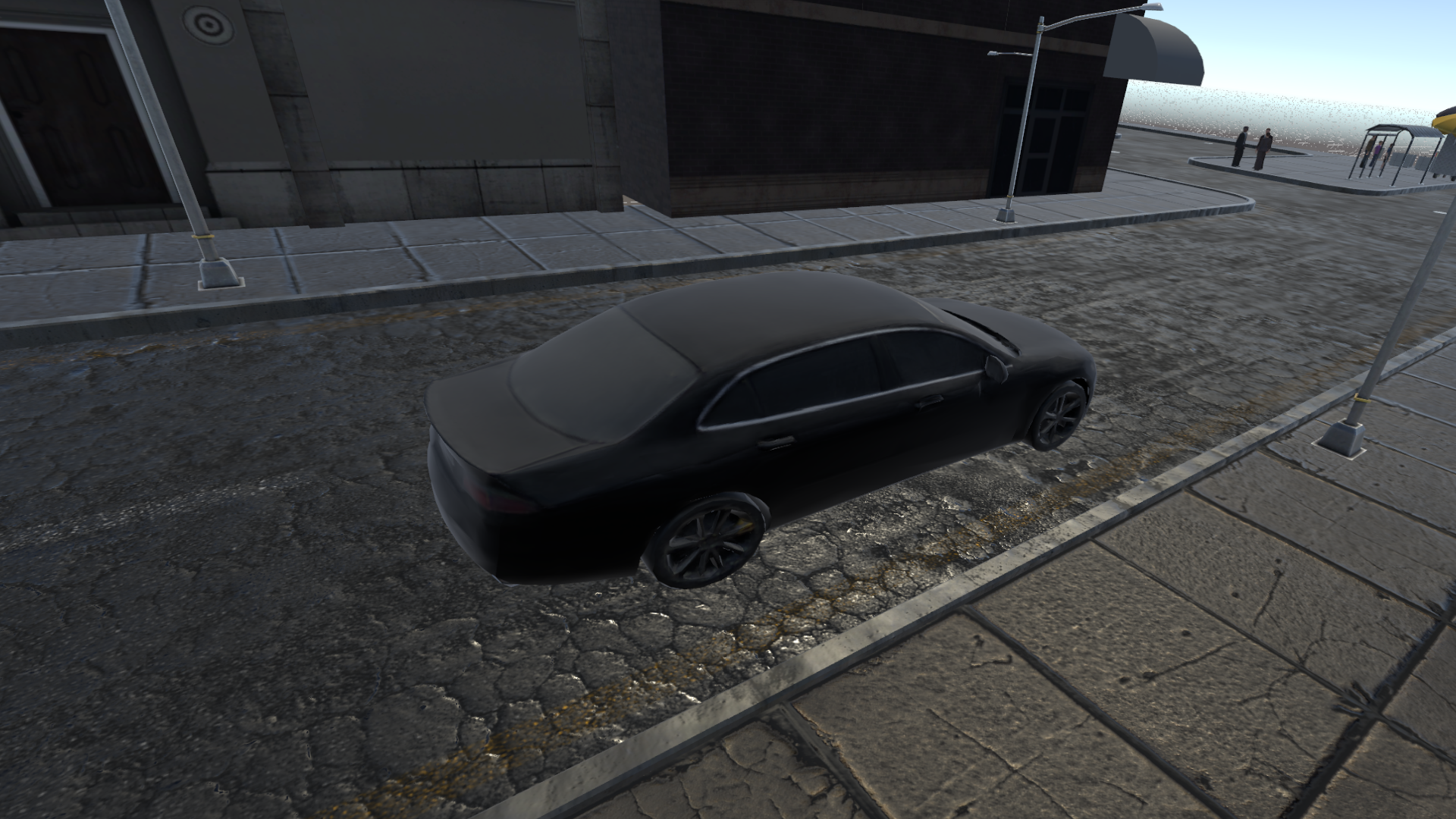}
        \caption{Scene after the prompt for asset generation..}
        \label{fig:assetgen2}
    \end{subfigure}

    \caption{Asset generation in XARA.}
    \label{fig:assetgen}
\end{figure}
The FastAPI server receives the text prompt and passes it to a pretrained TRELLIS-text-xlarge model. The model generates the requested asset and exports it as a GLB file. The current implementation is being developed to allow Unity to retrieve the generated file and spawn the asset directly into the scene in front of the user. Once integrated, the generated asset will be able to be selected, moved, rotated, scaled, and deleted like other scene objects.

Generation is also being integrated into the XR menu through a “Generate” interface, where the voice toggle feature (Section 3.4.4) will activate prompt input. The implementation will allow users to regenerate an asset if the result is unsatisfactory or save it for later use. Saved assets will be stored in the server directory and listed in the “Generated Assets” menu. The planned scene persistence functionality will record each asset's server path alongside its transform, allowing generated assets to be reloaded with the scene and shared between sessions and collaborators.

\section{Discussion}
\label{S:discussion}

XARA addresses a central challenge in XR-based transportation research: the gap between the domain expertise required to design meaningful behavioural experiments and the technical expertise required to implement them in immersive environments. By introducing natural language as a direct interface for scene manipulation, XARA allows researchers to act as authors of the experimental environment rather than as requirements providers dependent on technical intermediaries.

The integration of LLMs with XR experiment creation offers several advantages over conventional workflows. The most immediate advantage is accessibility. By replacing game-engine programming with voice-based natural language interaction, XARA lowers the technical barrier for XR experiment creation. Transportation researchers with limited programming experience can construct and modify experimental scenarios independently, which may broaden the use of immersive methods across behavioural travel research, safety analysis, human factors studies, and policy testing.

A second advantage is the speed of iteration. In conventional workflows, modifying an experimental scenario may require exiting the XR environment, editing the scene or code in Unity, and then re-entering the immersive setting, with additional build time required when application-level changes are involved. XARA allows many scene-level modifications to occur in real time through voice commands, substantially shortening the iteration cycle and supporting a more exploratory approach to experiment design, where researchers can adjust scenarios as hypotheses evolve.

A third advantage is the naturalness of voice interaction. Spoken language provides an intuitive way to describe spatial arrangements and actions. Commands such as ``put a car at the end of the road'' or ``move the pedestrian closer to the crossing'' align closely with how researchers conceptualize experimental scenes. This reduces the cognitive translation required to express the same intent through programming constructs, inspector panels, or graphical editor operations.

The dual-LLM architecture is also an important design choice. Separating speech recognition from command interpretation allows each component to specialize in a distinct task. Whisper provides acoustic robustness under variations in speaking rate, accent, microphone quality, and laboratory noise. GPT-4.1 provides semantic interpretation, mapping diverse natural language expressions to structured JSON commands. This separation also supports modularity: newer automatic speech recognition or natural language understanding models can replace existing components without requiring a redesign of the full system.

The current implementation is integrated directly within C\# and Unity, allowing voice-command processing, XR interaction, and scene manipulation to operate within the same application environment. External model requests are handled without blocking Unity's rendering loop, while interpreted commands are placed in a thread-safe queue for execution on the main Unity thread. This design helps preserve responsive XR interaction while maintaining a clear separation between language processing and scene execution.

Several limitations remain. The current implementation depends on network connectivity for API-based speech recognition and command interpretation, which introduces latency and dependence on external services. Although the heuristic fallback parser provides a graceful degradation path, its parsing capability is more limited than the LLM-based interpretation layer. The current command vocabulary is also focused on basic scene construction and object manipulation. More complex transportation-specific operations, such as traffic signal programming, vehicle trajectory definition, pedestrian behaviour scripting, and environmental condition control, require further development. Finally, XARA has not yet been formally evaluated with end users. Future usability studies are needed to compare voice-driven scene creation against conventional Unity-based workflows in terms of task completion time, usability, perceived workload, and researcher satisfaction.

\section{Future Work}
\label{S:future}

Future development of XARA can proceed along several directions. First, the system could be extended to support multi-user collaborative experiment design. A networked architecture would allow multiple researchers to occupy the same XR environment simultaneously and contribute to the construction of a shared experimental scenario. This would be particularly useful for interdisciplinary teams where traffic engineers, behavioural scientists, urban designers, and human factors researchers contribute different perspectives to scenario development. Such an extension would require synchronization of scene states across users, conflict-resolution mechanisms for simultaneous commands, and audio management to distinguish scene-directed voice commands from conversation between collaborators.

Second, XARA could be deployed as a cloud-based experiment infrastructure. A cloud deployment would allow researchers to access the system from different locations using XR headsets connected to a shared experiment platform. This model could reduce local hardware requirements, support distributed collaboration, and enable multi-site XR experiments involving participants in different geographic locations. It would also provide a route toward centralized model updates, version control, and shared experiment repositories.

Third, the system could be extended from command-level manipulation to full-scene generation from natural language descriptions. The current implementation processes individual commands that modify one object or attribute at a time. A future version could accept multi-sentence descriptions of complete scenarios. For example, a researcher could request a four-lane signalized intersection, several approaching vehicles, pedestrians waiting at the curb, dusk lighting, and wet road conditions in a single description. The LLM layer would then decompose the description into a sequence of executable commands while resolving spatial dependencies and maintaining physical plausibility.

Fourth, XARA could be integrated with automated behavioural data collection. Coupling the scene authoring system with eye tracking, physiological sensing, motion tracking, and response logging would create a more complete experiment lifecycle platform. Researchers could design the environment, expose participants to controlled scenarios, and automatically record gaze behaviour, movement trajectories, decision points, and response times. These behavioural data could be synchronized with the saved scene configuration, enabling more precise analysis of how specific environmental features influence participant behaviour.

Fifth, Asset Generation using TRELLIS will be hosted on a local server, fully integrated, and deployed into XARA. The goal of the Asset Generation feature is to enable users to create 3D assets dynamically based on their scene requirements. This allows users to access and generate assets directly within the XARA application, eliminating the need to source and import assets from external applications. 
\section{Conclusion}
\label{S:conclusion}

This paper presented XARA, an agentic dual-LLM system integrated with a Unity-based XR environment for voice-driven experiment generation in transportation research. The system addresses the technical barriers that currently limit the adoption of XR experiment methods by researchers without specialized game-engine development expertise.

XARA uses Whisper for automatic speech recognition and GPT-4.1 for natural language command interpretation. Spoken commands are translated into structured JSON instructions and executed in real time within the XR scene. The modular architecture, with Python handling language processing and C\# managing XR rendering and interaction, provides flexibility, robustness through fallback parsing, and a clear pathway for future component-level upgrades.

The system supports core scene manipulation operations, including object creation, movement, rotation, scaling, deletion, manual controller interaction, undo and redo operations, and scene persistence. These capabilities allow researchers to prototype and refine immersive transportation scenarios through an intuitive voice interface without directly editing Unity scenes or writing code.

By lowering the technical threshold for XR experiment creation, XARA has the potential to expand the use of immersive methods in transportation science. It enables a broader community of researchers to design controlled virtual environments for behavioural travel experiments, safety studies, human factors research, and policy testing. As language models, speech recognition systems, and XR hardware continue to mature, natural language interaction may become an increasingly important interface for designing and conducting virtual experiments.

In the future, we will implement a client-server architecture to enable the remote development of experiments and collection of the data. Currently, XARA only support development in Unity, but in the future we will enable the development in Unreal Engine as well. A systematic testing of the platform using a range of different users will be performed. We will also work towards developing a standardize protocol of XR experiments for travel behaviour.
\section*{Acknowledgements}
We would like to thank CFI for providing the equipment funding and we would like to thank our international collaborators (Vladimir Maksimenko, Robin Palmberg, Shun Su, Liting Yuan) for their feedback in the initial design stages of XARA.

\setlength{\bibsep}{0.0pt} 
\bibliographystyle{elsarticle-harv}
\bibliography{references}\newpage
\appendix

\end{document}